\documentclass[letterpaper,leqno]{IEEEtran}
\usepackage{mystyle}
\usepackage{algorithm2e}
\renewenvironment{proof}{\IEEEproof}{\endIEEEproof}
\newcommand{\mysize}{\scriptsize}
\newcommand{\SW}{\op{SW}}

\newcommand{\setBCO}{\rsfsB}
\newcommand{\setRCO}{\rsfsR}
\newcommand{\RCO}{R_{\op{CO}}}
\newcommand{\CSK}{C_{\op{SK}}}

\renewcommand{\odot}{\mathbin{
  \setbox0=\hbox{$\circ$}
  \cdot\kern-\wd0\kern+0.1em\ensuremath{\copy0}}
}

\title{On Tightness of Mutual Dependence Upperbound for
  Secret-key Capacity of Multiple Terminals}

\author{Chung Chan 
  \thanks{Manuscript written on May~20, 2008. Revised on Jun~15, 2008.}
  \thanks{Chung Chan
    (\href{mailto:chungc@mit.edu}{chungc@mit.edu}) is with the
    Laboratory of Information and Decision Systems, Department of
    Electrical Engineering and Computer Science, Massachusetts
    Institute of Technology.}  }

\date{May~20, 2008}

\begin{document}

\maketitle

\begin{abstract}
  Csisz\`ar and Narayan\cite{csiszar04} defined the notion of
  secret-key capacity for multiple terminals, characterized it as a
  linear program with Slepian-Wolf constraints of the related source
  coding problem of communication for omniscience, and upper bounded it
  by some information divergence expression from the joint to the product
  distribution of the private observations. 
  This paper proves that the bound is tight for the important case
  when all users are active, using the polymatroidal
  structure\cite{fujishige78} underlying the source coding problem.
  When some users are not active, the bound may not be tight. This
  paper gives a counter-example in which $3$ out of the $6$ terminals
  are active.
\end{abstract}

\section*{Acknowledgment}
I would like to thank Professor Lizhong Zheng for his guidance and
inspiring comments; Baris Nakiboglu for his construction of $\Ms_i$'s
in the proof of Theorem~\ref{thm:main} and for pointing out that
intersection are also tight constraints in Proposition~\ref{pro:pm};
Professor Imre Csisz\'ar for clarifications of the topic during his
visit at MIT; and, above all, my family for their support and
encouragements.

\section{Introduction}
\label{sec:introduction}

Consider a group of people with access to correlated discrete
memoryless sources. The problem of communication for
omniscience\cite{csiszar04} (CO) asks how much authenticated public
discussion is needed so that the active users in the group can recover
all the sources reliably. The problem of secret-key
generation\cite{csiszar04} (SK) asks how much secret they can agree
on.

Intuitively, the more correlated the sources are, the less public
discussion is needed for CO because active users can learn the sources
with their private observations as side information. With less
information revealed in public, the active users can also share more
secret. Are the maximum savings in public discussion and the amount of
secret exactly the mutual dependence of the sources? How should
one define such mutual dependence? 

In \cite{csiszar04}, Csisz\'ar and Narayan show that the secret-key
capacity is equal to the maximum savings in public discussion, i.e.
entropy rate of the sources minus the smallest rate of CO. While the
active users need not attain omniscience before agreeing on a secret
key, doing so guarantees certain universality result without loss of
optimality. They then upper bound the capacity by some normalized
information divergence\cite[(26)]{csiszar04} from the joint to the
product distribution, which is commonly interpreted as mutual
dependence of a set of random variables.\cite{fujishige78} As will be
shown in the sequel, this mutual dependence upperbound is indeed tight
for the case when all users are active. This gives an affirmative
answer to the first question, and confirms the information divergence
expression as a measure of mutual dependence. The equivalence is of
theoretical interest and can simplify the exact computation of
secret-key capacity in certain special cases as described in
\cite{csiszar04}.

In the other case when some users are helpers, it is straightforward
to show that the bound is tight for the $3$-terminal case by an
exhaustive test.\footnote{An exhaustive test by a computer program
  suggests informally that the bound is also tight for the
  $4$-terminal case. The $5$-terminal case is intractible.} We have
found, however, that the bound is lose for a $6$-terminal
counter-example with $3$ active users. It is unclear if there exists a
more general and ubiquitous mutual dependence expression that covers
this case and other related problems.

\section{Preliminaries}
\label{sec:preliminaries}

Let $\setM:=\Set{1,\dots,m}$ with $m\geq 2$ denote the set of all
terminals and $\RX_{\setM}:=(\RX_1,\dots,\RX_m)$ be the discrete
multiple memoryless sources (DMMS), which is a sequence of random variables
taking values from finite sets.  The subset
$\setA\subset\setM:\abs\setA\geq 2$ denotes the set of active users
while its complement $\setA^c$ is the set of helpers. From
\cite{csiszar04}, the set of (strongly) achievable rate tuples of CO
for $\setA$ is,
\begin{align}
  \setRCO(\setA) &= \Set{R_\setM\in `R^m: \forall
    \setB\in\setBCO(\setA),\;\SW(R_\setM,\setB)\geq 0}
  \label{eq:setRCO}
\end{align}
where
\begin{align}
  \setBCO(\setA) &:=\Set{\setB \subsetneq \setM:
    \setB\neq `0, \setB \nsupset \setA }\label{eq:setBCO}\\
  \SW(R_\setM,\setB) &:= \sum\nolimits_{j\in\setB} R_j - h(\setB)\label{eq:SW}\\
  h(\setB) &:= H(\RX_{\setB}|\RX_{\setB^c})\label{eq:h}
\end{align}
Each element in $\setBCO(\setA)$ corresponds to a Slepian-Wolf
constraint of CO for $\setA$. $\SW(R_\setM,\setB)$ is the constraint
function on the rate tuple $R_\setM:=(R_1,\dots,R_m)$. $h$ is the
conditional entropy function of the DMMS. The Slepian-Wolf constraint
is satisfied/tight/slack if $\SW(R_\setM,\setB)$ is $\geq$/$=$/$>$ zero.

From the set of achievable rate tuples, two key quantities of
interest, namely the smallest CO rate $\RCO(\setA)$ and secret-key
capacity $\CSK(\setA)$, can be computed as follows.
\begin{subequations}
  \label{eq:RCOCSK}
  \begin{align}
    \RCO(\setA)&= \min_{R_\setM\in \setRCO(\setA)} \sum_{i=1}^m R_i
    \label{eq:RCO}\\
    \CSK(\setA)&= h(\setM)-\RCO(\setA)\label{eq:CSK}
  \end{align}
\end{subequations}
Written in matrix form\cite{csiszar04}, $\RCO(\setA)$ is the solution to
the following primal and dual linear programs by the Duality
Theorem\cite[p.130-132]{dantzig97}.
\begin{subequations}
  \label{eq:LP}
  \begin{align}
    \RCO(\setA)
    &=\min_{\Mx :\MA\Mx\geq \Mb} \Mc\Mx && \text{(primal)}\label{eq:primal}\\
    &= \max_{\My\geq \M0: \My \MA = \Mc} \My\Mb && \text{(dual)}\label{eq:dual}
  \end{align}
\end{subequations}
where $\MA$ is an $l$-by-$m$ incidence matrix of the Slepian-Wolf
constraints;\footnote{$l$ is the total number $\abs{\setBCO(\setA)}$
  of constraints, which is $2^m-2^{m-\abs\setA}-1$.} $\Mb$ is the
corresponding $l$-by-$1$ column vector of conditional entropies; $\Mc$
is a $1$-by-$m$ row vector of all $1$'s; $\Mx$ is a $1$-by-$m$ rate
vector satisfying the Slepian-Wolf constraints $\MA\Mx\geq \Mb$; and
$\My$ is a $1$-by-$l$ non-negative weight vector.

Of particular interest is the following mutual dependence
upperbound\cite[(26)]{csiszar04} on $\CSK(\setA)$.
\begin{align}
  I(\setA) &:= \min_{2\leq k \leq
    \abs\setA,(\setC_1,\dots,\setC_k)\in\rsfsP_k(\setA)}
  I(\setC_1,\dots,\setC_k)
  \label{eq:I}
\end{align}
where
\begin{align}
  I(\setC_1,\dots,\setC_k)&:=
  \frac1{k-1} `1(\sum_{i=1}^k H(\RX_{\setC_i}) - H(\RX_{\setM})`2)
  \label{eq:I1}\\
  &= \frac1{k-1}
  D`1(\extendvert{P_{\RX_{\setM}} \| \prod\nolimits_{i=1}^k P_{\RX_{\setC_i}}}`2)\notag\\
  \rsfsP_k(\setA)&{
  \iftwocolumn
  \begin{multlined}[t]
    :=\big\{\Set{\setC_i}_{i=1}^k: \setC_1\cup\dots\cup\setC_k = \setM,\\
      \forall i\neq j,\;\setC_i \cap \setC_j = `0 , \setC_i\cap\setA
      \neq `0\big\}
  \end{multlined}
  \else
  :=\Set*{\Set{\setC_i}_{i=1}^k: \setC_1\cup\dots\cup\setC_k = \setM,
    \forall i\neq j,\;\setC_i \cap \setC_j = `0 , \setC_i\cap\setA
    \neq `0}
  \fi}
  \label{eq:P_k}
\end{align}
$\rsfsP_k(\setA)$ is the set of all $k$-partitionings of $\setM$ such
that each partition intersects $\setA$.  $I(\setC_1,\dots,\setC_k)$ is
the mutual dependence with respect to the $k$-partitioning
$(\setC_1,\dots,\setC_k)$.  $D(\cdot\|\cdot)$ is the information
divergence from the joint to the product
distribution.\cite{cover91eit} It is the well-known shannon's mutual
information in the $2$-terminal case, and therefore commonly
interpreted as mutual dependence for the multi-terminal
case.\cite{fujishige78} The goal of this paper is to confirm this
heuristic interpretation by studying the tightness of the mutual
dependence upperbound.

\section{Statement of results}
\label{sec:results}

From the derivation of the mutual dependence upperbound~\eqref{eq:I},
it is straightforward to see that tightness of the bound requires
certain Slepian-Wolf constraints to be tight. The necessary and
sufficient condition is as follows.

\begin{pro}[Tightness condition]
  \label{pro:tightness}
  $\CSK(\setA)=I(\setA)$ iff there exists $(\setC_1,\dots,\setC_k)\in
  \rsfsP_k(\setA)$ with $2\leq k \leq \abs\setA$ and $R_{\setM} \in
  \setRCO(\setA)$ such that $\SW(R_{\setM},\setC_i^c)=0$ for all $i\in
  \Set{1,\dots,k}$.
\end{pro}

Not only is this condition useful for a general exhaustive test for
tightness, it is also fundamental for the subsequent proof of the main
theorem. The idea is to look for a valid partitioning
$(\setC_1,\dots,\setC_k)$ such that
$\CSK(\setA)=I(\setC_1,\dots,\setC_k)$. This happens iff $\setC_i^c$'s
are tight Slepian-Wolf constraints.

Now, suppose $\setB_1$ and $\setB_2$ are two tight Slepian-Wolf
constraints. Are their union $\setB_1 \cup\setB_2$ and
intersection\footnote{The tightness of the intersection is not
  essential in the proof of the main theorem.} $\setB_1\cap\setB_2$
tight? The answer is affirmative as stated below due to the
polymatroidal structure\cite{fujishige78}.

\begin{pro}[Polymatroidal structure]
  \label{pro:pm}
  For all $R_{\setM}\in \setRCO(\setA)$ and
  $\setB_1\cup\setB_2\in\setBCO(\setA)$, we have
  \begin{align*}
    \left\{
    \begin{aligned}
      \SW(R_{\setM},\setB_1)&=0\\
      \SW(R_{\setM},\setB_2)&=0
    \end{aligned}\right.
    \implies
    \left\{
    \begin{aligned}
      \SW(R_{\setM},\setB_1\cup\setB_2)&=0\\
      \SW(R_{\setM},\setB_1\cap\setB_2)&=0
    \end{aligned}\right.
  \end{align*}
\end{pro}

When all users are active, the Duality Theorem and the induced
tightness of the union of tight constraints implies the existence of
the desired partitioning for tightness. This, however, does not extend
to the case when some users are helpers. The bound is indeed loose for
a particular $6$-terminal counter-example. Hence, we have the
following main theorem.

\begin{thm}[Tightness]
  \label{thm:main}
  When all users are active, we have 
  \begin{align*}
    \CSK(\setM) &= I(\setM)
  \end{align*}
  When some users are helpers, i.e.\ $\setA\subsetneq\setM$, there is
  a counter-example for which 
  $\CSK(\setA)< I(\setA)$ with strict inequality. 
\end{thm}

\section{Interpretation of mutual dependence}
\label{sec:interpretation}

Why is $I(\setM)$ a measure of mutual dependence? In the $2$-terminal
case, it is simply the well-known Shannon's mutual information $I(\RX_1; \RX_2):=D(P_{\RX_1\RX_2}\|P_{\RX_1}\circ
P_{\RX_2})$.\cite{cover91eit} But in the multi-terminal case, how can we
interpret the normalization factor and the minimization over different
partitionings?

Consider the sunflower example where
$\RX_i=(\RY,\RZ_i)$ for $i\in\setM$. The core $\RY$ and the petals
$\RZ_i$'s are mutually independent. For any
$k$-partitioning $(\setC_1,\dots,\setC_k)$,
\begin{align*}
  I(\setC_1,\dots,\setC_k)
  &= \frac1{k-1} `1( \sum_{i=1}^k H(\RX_{\setC_i}) - H(\RX_{\setM})
  `2)
  = H(\RY)
\end{align*}
since $H(\RX_{\setC})=H(\RY)+\sum_{j\in \setC} H(\RZ_{j})$ for
$\setC=\setM,\setC_1,\dots,\setC_k$. Thus, $I(\setM)=H(\RY)$ as
expected. From this example, we see that the factor $1/(k-1)$
compensates for the over-counting of $H(\RY)$ in the sum $\sum_{i=1}^k
H(\RX_{\setC_i})$, and that the optimal partitioning need not be
unique.

Suppose $(\setC_1,\dots,\setC_k)$ is an optimal partitioning that
achieves the minimum $I(\setM)$. Can we say that the random variables
within the same partition are more correlated than those in different
partitions? With the tightness result, we can recast the question in
the CO setting as follows. Consider grouping users according to an
optimal partitioning. If there is a private discussion that leads to
omniscience within each group before the public discussion, i.e.\ user
$j$ in $\setC_i$ knows the conglomerated sequence of $\RX_{\setC_i}$,
does it reduce the smallest CO rate? No because
$(\Set{1},\dots,\Set{k})$ is a valid partitioning that achieves the
same mutual dependence of the set of conglomerated random variables.
This means that attaining omniscience across different groups is the
bottleneck of the CO problem. One does not need addition redundancy in
rate to attain omniscience within each group. i.e.\
$\SW(R_{\setM},\setC_i^c)=0$ for all $i\in\Set{1,\dots,k}$ and some
optimal $R_{\setM}\in \setRCO(\setM)$. This gives an operational
meaning of the optimal partitioning.

\section{Counter-example}
\label{sec:counter-example}

In this section, we describe a counter-example for which the mutual
dependence upper bound $\CSK(\setA)\leq I(\setA)$ is loose, i.e.
satisfied with strict inequality, when some users are helpers.  Let
$m=6$, $\setA=\Set{1,2,3}$ and $(\RY_1,\RY_2,\RY_3,\RY_4)$ be iid
uniformly random bits. The DMMS $\RX_{\setM}$ is defined below as
the XOR $\oplus$ of every distinct pair of the random bits.
\begin{align*}
  \begin{matrix*}[r]
    \RX_1 := & \RY_1 & & \oplus \RY_3 & \\
    \RX_2 := & \RY_1 & & & \oplus \RY_4 \\
    \RX_3 := & & & \RY_3 & \oplus \RY_4 \\
    \RX_4 := & & \RY_2 & \oplus \RY_3 & \\
    \RX_5 := & & \RY_2 & & \oplus \RY_4 \\
    \RX_6 := & \RY_1 & \oplus \RY_2 & &
  \end{matrix*}
\end{align*}
There are altogether $l=55$ Slepian-Wolf constraints
$\setB\in\setBCO(\setA)$. With some algebra, the conditional entropy
function $h$ can be simplified to the following expression that depends
only on the cardinality of the constraint.
\begin{align}
  h(\setB)=
  \begin{cases}
    0 & \text{if $\abs\setB\in\Set{1,2}$}\\
    1 & \text{if $\abs\setB\in\Set{3,4}$}\\
    2 & \text{if $\abs\setB=5$}
  \end{cases}
  \label{eq:3}
\end{align}
Solving the linear program~\eqref{eq:RCO} and applying \eqref{eq:CSK},
\begin{align*}
  \RCO(\setA)=\frac{9}4 \quad \implies \quad
  \CSK(\setA)=1.75
\end{align*}
which is achieved by the unique optimal rate tuple with 
\begin{align*}
  R_1&=R_2=R_3=\frac14 &
  R_4&=R_5=R_6=\frac12
\end{align*}
Appendix~\ref{sec:matlab} describes how to compute and verify this
solution.

It remains to show that $I(\setA)>1.75$. Consider a $3$-partitioning
$(\setC_1,\setC_2,\setC_3)\in\rsfsP_3(\setA)$ such that each partition
has $2$ elements. Applying \eqref{eq:3} with $\abs{\setC_i^c}=4$,
\begin{align*}
  I(\setC_1,\setC_2,\setC_3)
  &= h(\setM)-\frac{h(\setC_1^c)+h(\setC_2^c)+h(\setC_3^c)}{3-1}\\
  &= 4-\frac{1+1+1}{2}=2.5
\end{align*}
All other cases can be computed analogously as follows by enumerating the
integer partitionings of $6$. 
{\renewcommand{\arraystretch}{1.3}
\begin{center}
\begin{tabular}{ccc}
  integer partitioning & $\Set{\abs{\setC_1^c},\dots,\abs{\setC_k^c}}$
  & $I(\setC_1,\dots,\setC_k)$\\\hline
  $2+2+2$ & $\Set{4,4,4}$ & $4-\frac{1+1+1}2=2.5$\\
  $3+2+1$ & $\Set{3,4,5}$ & $4-\frac{1+1+2}{2}=2$\\
  $4+1+1$ & $\Set{2,5,5}$ & $4-\frac{0+2+2}{2}=2$\\
  $3+3$ & $\Set{3,3}$ & $4-(1+1)=2$\\
  $4+2$ & $\Set{2,4}$ & $4-(0+1)=3$\\
  $5+1$ & $\Set{1,5}$ & $4-(0+2)=2$
\end{tabular}
\end{center}
}
We have the desired strict inequality that,
\begin{align*}
  I(\setA)
  &=2>1.75=\CSK(\setA)
\end{align*}
Appendix~\ref{sec:matlab} gives an alternative argument that
explains how this counter-example is constructed.

\section{Proofs of results}
\label{sec:proofs}

The mutual dependence upperbound is derived in \cite{csiszar04} by
removing certain Slepian-Wolf constraints as follows.

\begin{pro}[Mutual dependence upperbound]
  \label{pro:bound}
  \begin{align}
    \label{eq:bound}
    \CSK(\setA) &\leq I(\setA) \qquad
    \forall \setA\subset \setM:\abs\setA\geq 2
  \end{align}
\end{pro}

\begin{proof}
  Consider a partitioning $(\setC_1,\dots,\setC_k)\in \rsfsP_k(\setA)$.
  \begin{align}
    \sum_{i=1}^k \sum_{j\in \setC_i^c} R_j 
    &= \sum_{i=1}^k `1[\sum_{j\in\setM} R_j - \sum_{j\in \setC_i}
    R_j`2]\notag\\
    &= (k-1) \sum_{j\in \setM} R_j\notag\\
    \therefore
    \sum_{j\in\setM} R_j 
    &= \frac1{k-1} \sum_{i=1}^k \sum_{j\in \setC_i^c} R_j\label{eq:1}
  \end{align}
  Applying only the Slepian-Wolf constraints $\setC_i^c$'s on
  $R_{\setM}$ to the R.H.S.\ gives the desired bound.
\end{proof}

The tightness condition then follows from the tightness of those
Slepian-Wolf constraints used to derive the bound.

\begin{proof}[Proof of Proposition~\ref{pro:tightness}]
  Consider proving the `if' case. Let $(\setC_1,\dots,\setC_k)\in
  \rsfsP_k(\setA)$ and $R_{\setM} \in \setRCO(\setA)$ be such that
  $\SW(R_{\setM},\setC_i^c)=0$ for all $i\in\Set{1,\dots,k}$. Then, from
  \eqref{eq:1}, 
  \begin{align*}
    \sum_{j=1}^m R_j 
    &= h(\setM)-I(\setC_1,\dots,\setC_k)\\
    &\leq h(\setM)-I(\setA)
  \end{align*}
  By definition $\RCO(\setA)\leq \sum_{i=1}^m R_i\leq
  h(\setM)-I(\setA)$. Together with \eqref{eq:bound}, we have
  $\RCO(\setA)=H(\RX_{\setM})-I(\setA)$.

  Consider proving the contrapositive of the `only if' case. Suppose
  for all $2\leq k\leq \abs\setA$,
  $(\setC_1,\dots,\setC_k)\in\rsfsP_k(\setA)$, $R_{\setM}\in
  \setRCO(\setA)$, there exists $i\in\Set{1,\dots,k}$ such that
  $\SW(R_{\setM},\setC_i^c)>0$. Then, for all optimal $R_{\setM}$,
  \begin{align*}
    \RCO(\setA)&= \sum_{j=1}^m R_j> H(\RX_{\setM}) - I(\setA)
  \end{align*}
  which implies $\RCO(\setA)\neq H(\RX_{\setM})-I(\setA)$ as desired.
\end{proof}

The polymatroidal structure\cite{fujishige78} of the source coding
problem can be used to prove that union and intersection of tight
Slepian-Wolf constraints are tight as follows.

\begin{proof}[Proof of Proposition~\ref{pro:pm}]
    Consider some $R_{\setM}\in \setRCO(\setA)$ and
  $\setB_1\cup\setB_2\in\setBCO(\setA)$ such that
  $\SW(R_{\setM},\setB_1)=0$ and $\SW(R_{\setM},\setB_2)=0$.  Then,
  $\setB_1\cup\setB_2 \in \setBCO(\setA)$ implies that
  $\setB_1\cup\setB_2$, $\setB_1$, $\setB_2$ and $\setB_1\cap \setB_2$
  do not contains $\setA$. If they are non-empty, they are all in
  $\setBCO(\setA)$ by definition. Let $r(\setB):=\sum_{i\in\setB} R_i$
  for all $\setB\subset \setM$. Then,
  \begin{align*}
    r(\setB_1) &= h(\setB_1) & 
    r(\setB_1 \cup \setB_2) &\geq h(\setB_1 \cup \setB_2)\\
    r(\setB_2) &= h(\setB_2) &
    r(\setB_1 \cap \setB_2) &\geq h(\setB_1 \cap \setB_2)
  \end{align*}
  To show that the union is a tight constraint,
  \begin{align*}
    h(\setB_1 \cup \setB_2) \leq 
    r(\setB_1 \cup \setB_2)
    &= r(\setB_1) + r(\setB_2) - r(\setB_1 \cap \setB_2)\\
    &\leq h(\setB_1) + h(\setB_2) - h(\setB_1 \cap \setB_2)\\
    &\leq h(\setB_1 \cup \setB_2)
  \end{align*}
  where the last inequality is by the supermodularity
  of $h$. (see Lemma~\ref{lem:supermodularity} in
  Appendix~\ref{sec:clarifications}) Hence, $r(\setB_1 \cup
  \setB_2)=h(\setB_1 \cup \setB_2)$. Similarly, $r(\setB_1 \cap
  \setB_2)=h(\setB_1 \cap \setB_2)$. The case when $\setB_1 \cap
  \setB_2=`0$ holds trivially.
\end{proof}

The idea of the proof for the main theorem is to first obtain an
initial set of tight Slepian-Wolf constraints from the dual linear
program. Then, use the fact that union of tight Slepian-Wolf
constraints are tight to construct the desired partitioning for the
tightness condition.

\begin{proof}[Proof of Theorem~\ref{thm:main}]
  Consider some optimal solution $\My$ to the dual linear
  program~\eqref{eq:dual} with $\setA=\setM$. Let $t>1$ be the number
  of non-zero entries in $\My$. Construct the $1$-by-$t$ row vector
  $\tMy$ by eliminating the zero entries in $\My$; and the $t$-by-$m$
  submatrix $\tMA$ of $\MA$ by eliminating the corresponding rows,
  i.e. removing the $i$-th row if $y_i=0$ for $i\in\Set{1,\dots,l}$.

  Let $\tMA_{\bullet j}$ be the $j$-th column of $\tMA$. Construct the
  desired partitioning $(\setC_1,\dots,\setC_k)\in\rsfsP_k(\setM)$ by
  partitioning $\tMA$ into classes of identical columns and let the
  $t$-by-$1$ vector $\Ms_i^{\opT}$ be the column of the $i$-th class.
  The precise construction is as follows.

  \begin{algorithm}[H]
    \SetLine
    \KwIn{$\tMA$}
    \KwOut{$k$, $(\Ms_1,\dots,\Ms_k)$ and $(\setC_1,\dots,\setC_k)$}
    $k:=1$;\quad
    $\setC_1:=`0$;\quad
    $\Ms_1:=\tMA_{\bullet 1}^{\opT}$\;
    \For{$j:=1$ to $m$}{
      \uIf{$\tMA_{\bullet j}= \Ms_i^{\opT}$ for some $i\leq k$}{
        Add $j$ to $\setC_i$\;}
      \Else{
        $k:=k+1$;\quad
        $\setC_k:=`0$;\quad
        $\Ms_k:=\tMA_{\bullet j}^{\opT}$\;
      }
    }
  \end{algorithm}

  To argue that $k\geq 2$, note that $\tMA$ does not have any rows of
  all $1$'s nor rows of all $0$'s because $\setM$ and $`0$ are not
  Slepian-Wolf constraints. With $t>1$, at least two columns of $\tMA$
  are distinct. The other constraint that $k\leq m$ holds trivially.

  It remains to argue that $\setC_i^c$'s are tight Slepian-Wolf
  constraints in essence of the tightness condition in
  Proposition~\ref{pro:tightness}. As an immediate consequence of the
  Duality Theorem (see Lemma~\ref{lem:complementary} in
  Appendix~\ref{sec:clarifications}), rows of $\tMA$ correspond to
  tight Slepian-Wolf constraints. Since unions of tight constraints are
  tight by the polymatroidal structure (see Proposition~\ref{pro:pm}),
  it suffices to show that $\setC_i^c$'s are unions of constraints
  corresponding to rows of $\tMA$. 

  To do so, define $\neg$ as elementwise negation and $\odot$ as the
  logical matrix multiplication, in which addition and multiplication
  are replaced by logical `or' and `and'. Then, $(\neg \Ms_i) \odot
  \tMA$ corresponds to taking union of constraints in $\tMA$ whose
  corresponding entry in $\Ms_i$ is $0$. It suffices to show that
  $\Ms_i^c \odot \tMA$ indeed corresponds to $\setC_i^c$, or
  equivalently, that $(\neg\Ms_{i'}) \odot \Ms_i^{\opT}=1$.

  Assume to the contrary that there exists $i\neq i'$ in
  $\Set{1,\dots,k}$ such that $(\neg\Ms_{i'}) \odot \Ms_i^{\opT}=0$
  instead.  Since $\Ms_i^{\opT}$'s are constructed from distinct
  columns of $\tMA$, there exists $j\neq j'$ in $\setM$ such that,
  \begin{align*}
    (\neg\tMA_{\bullet j'})^{\opT} \odot \tMA_{\bullet j} =0 
    \quad \text{but} \quad
    \tMA_{\bullet j'}\neq \tMA_{\bullet j}
  \end{align*}
  This implies that the constraint of $\tMA_{\bullet j'}$ is a proper
  superset of that of $\tMA_{\bullet j}$. Since $\tMy>0$,
  \begin{align*}
    \tMy \tMA_{\bullet j'} &> \tMy \tMA_{\bullet j}
  \end{align*}
  But this contradicts $\tMy\tMA=\My\MA=\M1$, which is the constraint
  of the dual linear program. This completes the proof. (See
  Section~\ref{sec:counter-example} for the counter-example.)
\end{proof}

\section{Conclusion}
\label{sec:conclusion}

The mutual dependence upperbound $I(\setM)$ on the secret-key capacity
in \cite{csiszar04} is proved to be tight for the case when all users
are active. This gives an operational meaning to the mutual dependence
expression, and therefore confirms its heuristic interpretation as a
measure of correlations among a set of random variables.

The proposed proof uses the polymatroidal structure in the source coding
problem pointed out by \cite{fujishige78}. Starting with an
arbitrary solution to the dual problem (by the Duality Theorem in
linear programming already mentioned in \cite{csiszar04}), an initial
set of tight Slepian-Wolf constraints is obtained. The desired set of
tight Slepian-Wolf constraints is then derived using the
polymatroidal structure.

As shown by the counter-example with $3$ active users and $3$ helpers,
the mutual dependence upperbound need not be tight for the case when
some users are helpers. Thus, the mutual dependence expression $I(\setA)$
in this case is not supported with the operational meaning of CO and
SK, even though an exhaustive test shows that it is tight for the $3$-terminal
case and (informally with the help of a computer) for the $4$-terminal
case. It is unclear if there is a more general mutual dependence
expression that cover this case or other problems related to
the mutual dependence of a set of random variables.

\appendices

\section{Clarifications}
\label{sec:clarifications}

\begin{lem}[Supermodularity]
  \label{lem:supermodularity}
  For all $\setB_1,\setB_2\subset \setM$,
  \begin{align}
    h(\setB_1)+h(\setB_2) 
    &\leq h(\setB_1 \cup \setB_2) + h(\setB_1 \cap \setB_2)
    \label{eq:supermodularity}
  \end{align}
\end{lem}

\begin{proof}
  Subtracting L.H.S.\ from R.H.S.\ gives,
  \begin{align*}
    &\quad h(\setB_1 \cup \setB_2) + h(\setB_1 \cap \setB_2)
    -h(\setB_1)-h(\setB_2)\\
    &= -H(\RX_{\setB_1^c\cap\setB_2^c})-H(\RX_{\setB_1^c\cup\setB_2^c}) 
    +H(\RX_{\setB_1^c})+ H(\RX_{\setB_2^c})\\
    &= -H(\RX_{\setB_1^c\cup\setB_2^c}|\RX_{\setB_1^c\cap\setB_2^c}) 
    +H(\RX_{\setB_1^c}|\RX_{\setB_1^c\cap\setB_2^c})+
    H(\RX_{\setB_2^c}|\RX_{\setB_1^c\cap\setB_2^c})\\
    &= I(\RX_{\setB_1^c};\RX_{\setB_2^c} |
    \RX_{\setB_1^c \cap \setB_2^c})
  \end{align*}
  which is positive as desired.
\end{proof}

\begin{lem}[Complementary slackness{\cite[p.135-136]{dantzig97}}]
  \label{lem:complementary}
  Consider the primal~\eqref{eq:primal} and dual~\eqref{eq:dual}
  formulations for $\RCO(\setA)$. For all optimal solutions $\Mx$ and
  $\My$, $i\in\Set{1,\dots,l}$, if $y_i>0$, then the $i$-th row
  $\MA_{i \bullet}$ of $\MA$ corresponds to a tight Slepian-Wolf
  constraint, i.e.\ $\MA_{i \bullet} \Mx
  = b_i$.
\end{lem}

\begin{proof}
  By the (strong) Duality Theorem, $\Mc\Mx=\My\Mb$. Since $\My\MA=\Mc$
  for $\My$ to be feasible, we have $\My\MA\Mx = \My\Mb$, or
  equivalently,
  \begin{align*}
    \sum_{j=1}^l y_j \MA_{j \bullet}\Mx &= \sum_{j=1}^l y_j b_j
  \end{align*}
  Assume to the contrary that there exists $i\in\Set{1,\dots,l}$ such that
  $y_i >0$ and $\MA_{i \bullet}\Mx \neq b_i$. Since $\MA\Mx\geq \Mb$,
  and $\My\geq\M0$, the L.H.S.\ of the last equation would instead be
  strictly larger than the R.H.S., which is a contradiction.
\end{proof}

\section{Computations for the counter-example}
\label{sec:matlab}
\newcommand{\mycode}{\lstinline[basicstyle=\small\ttfamily]}

\lstset{language=Matlab,morekeywords={linprog,mpt_init,mpt_solveLP}}

In this section, we compute $\RCO(\setA)$ for the counter-example in
Section~\ref{sec:counter-example} using the Multi-Parametric Toolbox\cite{mpt}
for Matlab. First, we initialize the toolbox to use an LP solver
that always terminates at a vertex optimal solution.

\begin{lstlisting}[linerange={1-2}]
% initialize the LP solver with CDD Criss-Cross
opt=mpt_init('lpsolver','cdd','abs_tol',1e-8);
\end{lstlisting}

The matrices for the linear program~\eqref{eq:primal}
can be constructed manually as follows.

\begin{lstlisting}[linerange={4-23},basicstyle=\mysize]
A=[
 1 0 0 0 0 0;0 1 0 0 0 0;1 1 0 0 0 0;0 0 1 0 0 0;1 0 1 0 0 0;
 0 1 1 0 0 0;0 0 0 1 0 0;1 0 0 1 0 0;0 1 0 1 0 0;1 1 0 1 0 0;
 0 0 1 1 0 0;1 0 1 1 0 0;0 1 1 1 0 0;0 0 0 0 1 0;1 0 0 0 1 0;
 0 1 0 0 1 0;1 1 0 0 1 0;0 0 1 0 1 0;1 0 1 0 1 0;0 1 1 0 1 0;
 0 0 0 1 1 0;1 0 0 1 1 0;0 1 0 1 1 0;1 1 0 1 1 0;0 0 1 1 1 0;
 1 0 1 1 1 0;0 1 1 1 1 0;0 0 0 0 0 1;1 0 0 0 0 1;0 1 0 0 0 1;
 1 1 0 0 0 1;0 0 1 0 0 1;1 0 1 0 0 1;0 1 1 0 0 1;0 0 0 1 0 1;
 1 0 0 1 0 1;0 1 0 1 0 1;1 1 0 1 0 1;0 0 1 1 0 1;1 0 1 1 0 1;
 0 1 1 1 0 1;0 0 0 0 1 1;1 0 0 0 1 1;0 1 0 0 1 1;1 1 0 0 1 1;
 0 0 1 0 1 1;1 0 1 0 1 1;0 1 1 0 1 1;0 0 0 1 1 1;1 0 0 1 1 1;
 0 1 0 1 1 1;1 1 0 1 1 1;0 0 1 1 1 1;1 0 1 1 1 1;0 1 1 1 1 1;
 ];

b=[
 0 0 0 0 0 0 0 0 0 0 0 1 0 0 0 0 0 0 0 1 0 0 0 1 0,...
 1 1 0 0 0 1 0 0 0 0 0 0 1 0 1 1 0 0 0 1 0 1 1 1 1 1 2 1 2 2
 ]';

c=[1 1 1 1 1 1];
\end{lstlisting}

We can then use the command \mycode!mpt_solveLP! to solve the primal and dual
linear programs. To match the API, \mycode!y!  is a column vector instead of
a row vector, and \mycode!nRCO! is the negation of $\RCO(\setA)$.

\begin{lstlisting}[linerange={25-31}]
% Primal: $Rco=\min_x c*x$    subject to:   $-A*x <= -b$ 
[x,Rco]=mpt_solveLP(c',-A,-b);

% Dual: $nRco=\min_y b'*y$    subject to:   $A'*y = c'$ 
lb=zeros(size(A,1),1);
ub=Inf*ones(size(A,1),1);
[y,nRco]=mpt_solveLP(-b,[],[],A',c',[],[],lb,ub);
\end{lstlisting}

The optimal solution \mycode!x==[.25 .25 .25 .5 .5 .5]! of the primal
achieves \mycode!Rco==2.25!. The optimal solution of the dual has non-zero
entries equal to $0.25$ at positions of the tight Slepian-Wolf
constraints. One can check that
\mycode!y(A*x-b<=opt.abs_tol)!
returns a sequence of all $0.25$'s, and 
\mycode!sum(A*x-b>opt.abs_tol)! 
is $0$. Now, to verify that the computed \mycode!x! is indeed optimal (since
the output from a computer cannot be used as a formal proof), one can
check that \mycode!x! and \mycode!y! are both feasible and satisfy the primal/dual
optimality criteria in Theorem~2.9 of \cite[p.48]{dantzig97b}. This
implies that both \mycode!x! and \mycode!y! are optimal, and $\RCO(\setA)=2.25$ as
desired. The mutual dependence upperbound computed in
Section~\ref{sec:counter-example} is therefore formally proved to be
loose.

There is an alternative explanation that the bound is loose without
calculating it explicitly. The underlying reasoning has guided the
construction of the counter-example, and the proof of tightness in the
case when all users are active. Consider the tightness condition in
Proposition~\ref{pro:tightness}. If one can show that for all optimal
solutions in $\setRCO(\setA)$, there is no subset of $2\leq k\leq
\abs\setA$ tight Slepian-Wolf constraints whose complement partitions
$\setM$, then the mutual dependence bound is loose. For the particular
optimal solution \mycode!x!, exactly six Slepian-Wolf constraints are tight.
i.e.
\begin{align*}
  \overbrace{
    \begin{bmatrix*}
      1&0&1&1&0&0\\
      0&1&1&0&1&0\\
      1&1&0&0&0&1\\
      1&1&0&1&1&1\\
      1&0&1&1&1&1\\
      0&1&1&1&1&1
    \end{bmatrix*}
  }^{\tMA:=}
  \overbrace{
    \begin{bmatrix*}
      \frac14\\ \frac14\\ \frac14\\ \frac12\\ \frac12\\ \frac12
    \end{bmatrix*}
  }^{\Mx:=}
  =
  \overbrace{
    \begin{bmatrix*}
      1\\1\\1\\2\\2\\2
    \end{bmatrix*}
  }^{\tMb:=}
\end{align*}
where $\Mx$ is the optimal solution \mycode!x!;
Rows of $\tMA$ corresponds to the tight Slepian-Wolf constraints; and
elements of $\tMb$ are the corresponding conditional entropies.
Note that rows of $\tMA$, has either one $0$ or three $0$'s.  For the
complements of a subset of $2$ or $3$ Slepian-Wolf constraints to
partition $\setM$, the only possibility is to have two tight
Slepian-Wolf constraints whose incidence vectors have three $0$'s. But
there is no two such rows of $\tMA$ having $0$'s at complementing
positions. Thus, the bound is loose if \mycode!x! is the unique optimal
solution, which can be shown using the PUFAS algorithm in
\cite{appa02} as follows.

We first express the primal linear program in equational form by
introducing a slack variable for each constraint to take up the slack.
Let $\Mx_s$ be the $55$-by-$1$ column vector of slack variables. Then,
\begin{align*}
  \RCO(\setA)=
  \min_{\MA \Mx \geq \Mb} \Mc \Mx
  &= \min_{\bSM \MA & -\MI\eSM \bSM \Mx \\ \Mx_s \eSM = \Mb,\bSM \Mx\\
    \Mx_s\eSM \geq 0} 
  \bM \Mc & \M0\eM \bM \Mx \\ \Mx_s \eM
\end{align*}
where $\MI$ is an identity matrix with matching dimensions. 
The vertex optimal solution \mycode!x1! of this equational form can be
computed as follows.

\begin{lstlisting}[linerange={33-38}]
% Solving the primal problem in equational form
ceq=[c zeros(1,size(A,1))];
Aeq=[A -eye(size(A,1))];
lbeq=zeros(size(Aeq,2),1);
ubeq=Inf*ones(size(Aeq,2),1);
[x1,Rco1]=mpt_solveLP(ceq',[],[],Aeq,b,[],[],lbeq,ubeq);
\end{lstlisting}

One can verify that \mycode!Rco1==2.25!, \mycode!x1(1:6)==x! and \mycode!x1(7:61)==A*x-b!
within the absolute tolerance. If the solution is not unique,
there will be a different vertex optimal solution with at least one
positive entry at the position where the corresponding element of
\mycode!x1! is zero. Let $\Md$ be the incidence row vector of $0$'s in
\mycode!x1!. Maximizing $\Md\Mx$ subject to an additional constraint for
optimality that $\bSM \Mc & \M0\eSM \bSM \Mx \\ \Mx_s \eSM$ equals \mycode!Rco1!
gives an alternative solution if there is one. If the
solution is unique, the optimal value would be $0$, achieved by
\mycode!x1!. This test can be implemented as follows.

\begin{lstlisting}[linerange={40-42}]
% Verifying uniqueness of optimal solution x (i.e. show notUnique=0)
[x2,notUnique]=mpt_solveLP(-double(x1<1e-8),[],[],...
  [Aeq; ceq],[b;Rco1],[],[],lbeq,ubeq);
\end{lstlisting}

One can check that \mycode!notUnique==0! and \mycode!x1==x2! within
the absolute tolerance. This shows \mycode!x! is the unique optimal
solution, and hence the mutual dependence bound is loose.

\bibliography{main}
\bibliographystyle{plainurl}

\end{document}